\documentclass[twocolumn,showpacs,preprintnumbers,amsmath,amssymb]{revtex4}

\usepackage{graphicx}
\usepackage{dcolumn}
\usepackage{bm}

\begin{document}


\title{Qubit dispersive readout scheme with a microstrip squid amplifier}

\author{S. Michotte$^{1,2}$}
\affiliation{$^{1}$Unité$\acute{e}$ de Physico-Chimie et de Physique des Mat$\acute{e}$riaux 
(PCPM), Universit$\acute{e}$ Catholique de Louvain, Place Croix du Sud 1, B-1348 Louvain-la-Neuve, Belgium
\\
$^{2}$Department of Physics, University of California, Berkeley
\\
Berkeley, CA 94720 USA
}

\date{\today}

\begin{abstract}

A flux qubit readout scheme involving a dispersive technique coupled to a microstrip squid amplifier has been tested experimentally. Thanks to the almost quantum limited noise of this amplifier at low temperature, this readout device is very promising for a design with actual qubits. It's proof of principle and low noise performance have been tested by simulating the qubit presence applying a tiny flux change in the input squid. The resonant frequency of the amplifier is adjusted in-situ with a varactor diode to approach the frequency of the tank circuit. Depending on the sign of the transfer function, two operating mode (crossing or anticrossing regime) have been identified.

\end{abstract}

\pacs{85.25.Cp ; 03.67.Lx ; 85.25.Dq}


\maketitle

Many systems such as Photons, ions, atoms, etc. are presently considered to implement a quantum computer, each of them with its own advantages and remaining challenges. Among these various possible building blocks lies one of the most macroscopic realization of a quantum bit or qubit that is a superconducting circuit. These electrical circuits such as flux qubit \cite{Intro1} can be viewed as engineered 'atoms' which could benefit from the advantage of lithography to scale up their number. Moreover, the coupling between two such qubits can be controlled and even switched off by the application of an appropriate superconducting current in a SQUID surrounding these qubits \cite{Travis1}. However, reducing the coupling to the environment that leads to decoherence is still the main challenge for these qubits. Indeed, T1 relaxation time and T2 dephasing time are currently too low to build a practical computer: they respectively amount to a few hundreds and a few tens of ns while about 10 $\mu s$ would be required. The difficulty arise from the fact that to stay coherent, the qubit has to be isolated the most from the rest of the world but in order to access it such as for the readout process, connections have to be established between it and the outside world, therefore bringing sources of decoherence. In our design, the same SQUID that is used to control the coupling is also used for the readout. In Ref.\cite{Travis1}, an unshunted DC SQUID switching readout was used with the disadvantage that the SQUID goes to the dissipative state. Due to this, repetition rates in relaxation-sensitive measurements may be no faster than 1 kHz, in order to allow for recombination of hot quasiparticles. Our current focus is to improve the readout process by using a dispersive readout technique that generates no heat and thus allows much faster repetition rates. Such a dispersive readout which measures a change in the dynamic inductance of the SQUID as it varies with the flux in the qubit loop enclosed has already been realized in ref. \cite{Readout1,Readout2,Readout3}. However, the noise temperature of the amplifier that is used for the readout is critical. Here, we demonstrate how to realize a dispersive readout with the microstrip squid amplifier (MSA)\cite{MSA1}. Indeed, in the 1 GHz range and at dilution fridge temperature, this amplifier is almost quantum limited \cite{MSA4} with a noise temperature of a few tens of mK, thus about 20 times better than today's best GaAs cryogenic amplifier which noise temperature cannot be made better than one Kelvin. Combining the MSA with a dispersive readout scheme potentially allows to decrease the measurement time of a qubit but also to couple less with the qubit to minimize the source of decoherence from the readout device.\\

\begin{figure}[hbtp]
\includegraphics[width=0.5\textwidth]{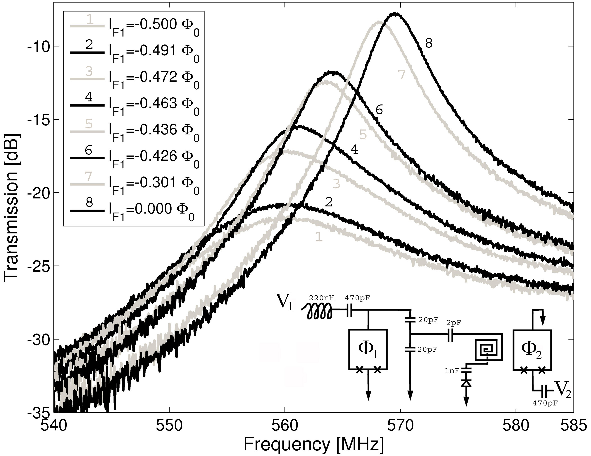}
\caption{\label{fig1} Transmission vs frequency curves obtained at 0.309K and with the MSA OFF, for different input flux in the tank circuit Squid ($I_{F1}$). After the attenuators, the incoming power is -99dBm. Inset is a schema of the low temperature part of the readout circuit.}
\end{figure}

To test experimentally the principle of a dispersive readout with the microstrip amplifier, we built (see inset of fig.\ref{fig1}) a tank circuit composed of an input squid and two 20 pF capacitors. The microwave readout pulse is send from the left (after the succession of a 49dB attenuator at ambient and a 30dB attenuator at 1K), into the 220nH inductor whose large impedance at microwave frequency avoid the quality factor of the tank to be loaded with the 50 Ohm line. The MSA is hook by taping the two tank capacitors with a coupling capacitor of 2pF. A 470pF capacitor is used on the right to ac couple the output $V_2$ of the microstrip amplifier (depicted here by a microstrip side by side with a second squid but in reality the microstrip lies on top of the squid washer with an insulating layer in between). Two HEMT post-amplifiers are subsequently used at 4K and at ambient with a total gain of 71dB. A varactor diode is place at the extremity of the microstrip in order to tune the frequency of the MSA \cite{MSA2}. DC current and flux bias lines (not shown) allow to bias the MSA at its optimum points ($I_{b2}$ and $\Phi_2$) as well as to apply the right flux in the input tank circuit squid ($\Phi_1$). 

Indeed, the input squid is the one supposed to enclose the qubit that has to be measured. A change in the qubit that couples its flux in this squid would lead to a change of the order of 10 $m\Phi_0$ in the squid where $\Phi_0=h/2e\approx2\cdot10^{-15}Tm^2$ is the flux quantum. In order to perform the readout in say 100ns, a flux noise referred to the input squid as low as 3 $\mu\Phi_0/\sqrt{Hz}$ has to be achieve. For this test of principle of the readout device, we didn't use a real qubit enclosed in the input squid but we rather verify that a change as small as 10 $m\Phi_0$ applied to it by the dc flux biasing coil leads to a good sensitivity $\Delta V_2/\Delta \Phi_1$ and the desired flux noise $\Phi_{1N}$. \\
Since the grounding of the washer of the MSA is not perfect (it is made via 5 small wirebounds in parallel), the microwave transmission although being low can still be detected if the MSA is turn OFF as can be seen on Fig \ref{fig1}. 
Here we only have a 10 MHz change in the resonant frequency $f_r$ of the tank circuit going from $\Phi_1=\Phi_0$ to $\Phi_0/2$ because the tank capacitors are located off-chip. This arrangement gives relatively large stray inductors due to the wirebounds and the surface mount capacitors. The resulting total
stray inductance $L_{stray}=8nH$ reduces the change of the resonant frequency roughly given by:
\begin{equation}
\label{fr}
f_r=\frac{1}{2\pi\sqrt{(L_{stray}+L_J(\Phi_1))\cdot C}}
\end{equation}
No dc bias current is applied to the tank circuit squid, so the phase difference $\delta(\Phi_1)$ is the same across each junction of the tank circuit squid and the josephson inducance (per junction) amounts to:
\begin{equation}
\label{LJ}
L_J(\Phi_1)=\frac{\Phi_0}{2\pi\cdot I_0 \cdot cos(\delta(\Phi_1))}
\end{equation}
where $I_0=4.5 \mu A$ at 0.3K is the junction zero field critical current and with
\begin{equation}
\label{delta}
\delta(\Phi_1)=\pi\frac{\Phi_1}{\Phi_0}+\frac{L_{self}}{2L_J^0}sin(\delta(\Phi_1))
\end{equation}
where $L_{self}\sim 15$ pH is the self inductance of the tank circuit squid and $L_J^0=L_J(\Phi_1=\Phi_0)=71$ pH at 0.3 K.
In order to have a good sensitivity, $\delta(\Phi_1)$ has to vary from 0 to almost $\pi/2$, which only happens if $L_{self}<<L_J^0$ as it is the case here.
As expected from equation \ref{LJ} and \ref{delta}, the change of the Josephson inductance (and thus also the one of the resonant frequency) is almost zero at $\Phi_1=\Phi_0$ and it increases strongly when approaching $\Phi_1=\Phi_0/2$. In order to have a good sensitivity as well as a good quality factor, we choose to work about $65m\Phi_0$ away from $\Phi_1=\Phi_0/2$, which corresponds i.e. on fig \ref{fig1} to curves 5 and 6.\\

\begin{figure}[hbtp]
\includegraphics[width=0.5\textwidth]{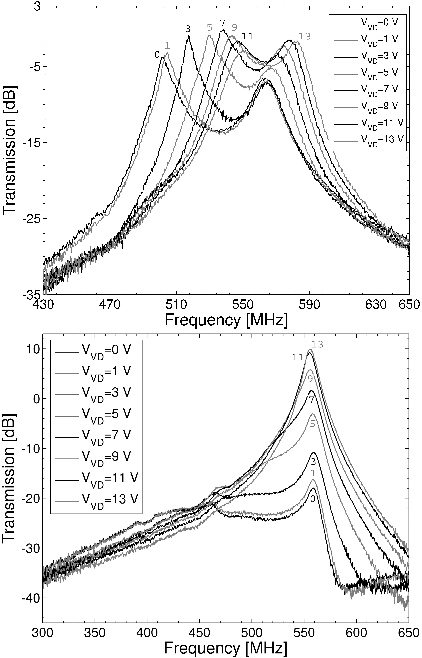}
\caption{\label{fig2} Transmission vs. frequency cures obtained at $\Phi_1=0.475\Phi_0$ with a network analyzer at 0.309K (-99dBm incoming power) with the MSA 'ON' i.e. optimally biased to have the largest respectively positive or negative transfer function (upper or lower figure). The voltage $V_{VD}$ across the varactor diode placed at the end of the microstrip is varied. Calibration was performed with the PC board illustrated in the inset of fig. 1 replaced by a short transmission line.}
\end{figure}

\begin{figure}[hbtp]
\includegraphics[width=0.5\textwidth]{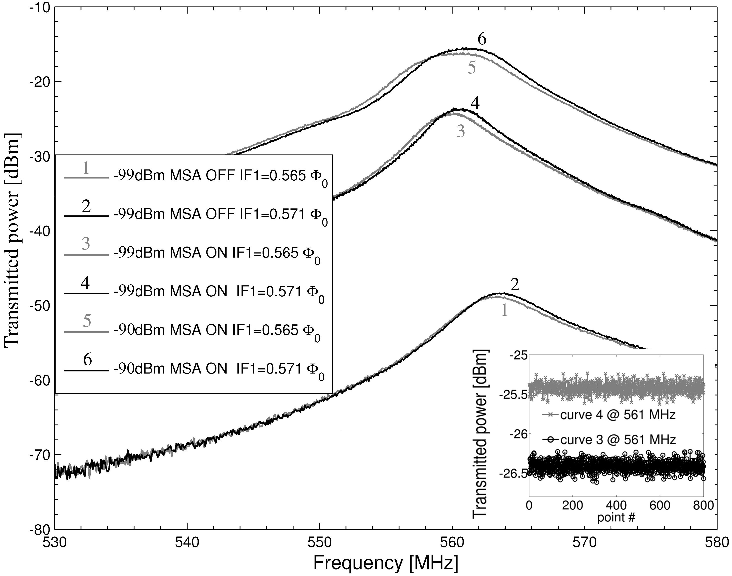}
\caption{\label{fig3} Transmitted power vs. frequency measured with a spectrum analyzer for a couple of input flux values (6m$\Phi_0$ apart) with the MSA OFF or ON with an incoming power of -99 or -90dBm. Inset: power spectum at 561MHz for curve 3 and 4.}
\end{figure}

We now turn 'ON' the microstrip amplifier i.e. we apply a dc current bias current $I_{b2}$ and a dc flux $\Phi_2$ such that the transfer function $dV_2/d\Phi_2$ and so the gain is maximized. Both optimum corresponding to the largest positive and negative transfer function (of respectively +600$\mu V/\Phi_0$ and -560$\mu V/\Phi_0$) have been considered (see fig. \ref{fig2}).
As we vary the voltage $V_{VD}$ across the varactor diode situated at the end of the microstrip, we can tune the frequency of the MSA. The flux in the tank circuit squid $\Phi_1$ is here equals to 0.475$\Phi_0$ (and so the tank frequency when the MSA is 'OFF' is about 560MHz) but the same behavior is seen even with $\Phi_1=\Phi_0$. If the positive transfer function is chosen, an anticrossing behavior is observed. Indeed, as we try to increase the frequency of the MSA to bring it closer to the one of the tank circuit, repulsion occurs and the peak in the transmission curve that was initially corresponding to the tank seems to be pushed away. Although our coupling capacitor is small, this behavior is the one expected given the high quality factor of the two resonators in play. An inductive readout was also performed in this anticrossing regime but the best results were obtained when the negative transfer function was chosen. In this case, instead of an avoided crossing, the two resonant frequencies allow a complete crossing when we try to put them on top of each other with even a small attraction between the two in this case. Such Different behaviors of the MSA according to the sign of its transfer function are related to feedback effects \cite{MSA3}.
As fig\ref{fig2} shows, the crossing case allows (see curve $V_{VD}=13V$) to have a gain about 10dB larger than in the anticrossing case, so we describe below the best readout performances obtained in this crossing case with $V_{VD}=13V$.\\
To perform a dispersive readout, since no energy can be left behind, the power send should be sufficiently low so that the critical current of the tank circuit squid is not reached. This is indeed the case if the incoming power is -99dBm. In this case, as can be seen on fig.\ref{fig3} (compare e.g. curves 1 and 3), the gain of the MSA is about 25dB. 
Curves 3 and 4 illustrate a $6m\Phi_0$ change applied at the input of this readout device (corresponding to a Josephson inductance change of 13pH per junction).
At the optimum readout frequency of 561MHz, the sensitivity referred to the output of the MSA is 69 $\mu V/\Phi_0$ and the flux noise referred to the input tank circuit squid is $\Phi_{1N}=4 \mu\Phi_0/\sqrt{Hz}$. An even better flux noise of $\Phi_{1N}=2 \mu\Phi_0/\sqrt{Hz}$ was achieved with an incoming power of -90dBm (and at a frequency of 562MHz) but then, the tank squid critical current was exceeded which is not desirable to measure a qubit in a dispersive way. The best flux noise obtained with this device at 0.3K is thus $\Phi_{1N}=4 \mu\Phi_0/\sqrt{Hz}$, which is very satisfactory. In the real device that would have to read a qubit, the two tank circuit capacitors will be on-chip, therefore reducing the stray inductors and thus producing an even better flux noise via an increase in the sensitivity. Moreover, going from 0.3K to 30mK will also reduce further the noise from the tank circuit as well as the noise from the MSA. Both these improvements would allow an even faster measurement of the qubit state or an even stronger decoupling of the qubit from this readout device (via a decrease of the coupling capacitor).
\\
In conclusion, we showed experimentally the feasibility of an inductive qubit readout scheme coupled to a microstrip squid amplifier. Satisfactory flux noise was already achieved at 0.3K with off-chip capacitors for the tank circuit. Even better performance could thus be achieved in the final design were capacitors will be on-chip and at an order of magnitude lower temperature. Such reading device involving the MSA as a first post-amplifier is a way to elude
one of the dominant noise source coming from the HEMT amplifier, which noise temperature is typically 2-4K.

\begin{acknowledgments}
The author whish to thank J. Clarke and D. Kinion for their help and support.
This work was supported by the NSF under grant EIA-020-5641, Air
Force Office of Scientific Research under grant F49-620-
02-1-0295, Army Research Office under grant DAAD-19-
02-1-0187, Advanced Research and Development
Activity, and Bavaria California Technology Center.
S. M. is a Postdoctoral Researcher of the National Fund for Scientific Research (FNRS) Belgium.
\end{acknowledgments}


\end{document}